**Variable Spectral Slope and Nonequilibrium Surface Wave Spectrum**

Paul A. Hwang[*]

Remote Sensing Division, Naval Research Laboratory, Washington, DC, 20375, USA

*Corresponding author*

*Email address:* paul.hwang@nrl.navy.mil

**Abstract**

The wave spectral properties in the centimeter to decameter (cmDm) wavelength range is of great interest to ocean remote sensing and studies of ocean surface processes including the surface roughness, air-sea energy and momentum exchanges, wave breaking, and whitecap coverage. For more than six decades, the cmDm wave components are generally considered to be in the equilibrium range, and its spectral function has a constant slope: -5 or -4 in the 1D frequency spectrum, and -3 or -2.5 in the 1D wavenumber spectrum. Some variations of the equilibrium spectrum models include varying the frequency spectral slope from -4 to -5 at some multiple of the spectral peak frequency, or incorporating a threshold velocity in the reference wind speed. Extensive efforts are then devoted to quantifying the spectral coefficient of the equilibrium spectrum function. The observed wind wave spectral slopes in the ocean environment, however, are rarely constant. The variable spectral slope is indicative of the nonequilibrium nature of surface wind waves in the field. As a result the wave properties in field observations are significantly different from those inferred from assuming a constant spectral slope. From signal-to-noise consideration the surface slope measurements are much more sensitive than the elevation data for the study of cmDm waves. Recently, several large datasets of low-pass-filtered mean square slope (LPMSS) have been reported in support of the Global Navigation Satellite System Reflectometry (GNSSR) tropical cyclone wind sensing effort. Combining the LPMSS observations with a spectrum model that accommodates a variable spectral slope, this paper seeks to quantify the connection between the variable spectral slope and the spectral properties of cmDm waves.

**1. Introduction**

Ocean surface roughness and surface wave breaking are two processes with significant contributions to microwave remote sensing of the ocean environment. Surface waves much shorter than the wavelength at the energy spectral peak are the major contributor of the ocean surface roughness. The dimensions of surface wave breaking patches are also one to two orders of magnitude shorter than the spectral peak wavelength. There are experimental evidences that these length scales of breaking patches and associated effects on ocean surface processes are detectable in the analyses of surface waveform time series, acoustic noise, radar sea spikes, whitecaps, surface wave spectrum, sea spray aerosol, or microwave sea surface brightness temperature (e.g., Hwang et al., 1989; Ding and Farmer, 1994; Frasier et al., 1998; Phillips et al., 2001; Melville and Matusov, 2002; Hwang and Wang, 2004; Hwang, 2007; Hwang et al., 2008, 2016; and references therein). Understanding of the short wave properties is very important to the study of a wide range of subjects including the ocean surface roughness and the related air-sea energy, momentum, and mass exchanges. The knowledge is also crucial to the interpretation of the ocean remote sensing signals using microwave techniques, as well as the design of ocean surface wave spectrum models. The spectral region of the short scale waves mentioned above has been considered to be in the equilibrium range, for which the papers by Phillips (1958, 1985) represent the pioneering works.

A key criterion of the equilibrium spectrum is the spectral slope. In Phillips (1958), the equilibrium or saturation frequency spectral slope is -5. The corresponding 1D wavenumber spectral slope is -3. This high frequency asymptote is adopted in the spectrum models of Pierson and Moskowitz (1964) and Hasselmann et al. (1973, 1976). The Phillips (1958) spectral function is used to study the Bragg resonance mechanism of radar backscattering in the early days (e.g., Wright, 1966, 1968).

Phillips (1985) presents an elegant source-function balance discussion and revises the equilibrium frequency spectral slope to -4. The corresponding 1D wavenumber spectral slope is -2.5. The -4 high frequency tail is incorporated in the wind wave spectrum model of Donelan et al. (1985). Some variations of the equilibrium spectrum modeling include: (a) prescribing the spectral slope to be -4 in a small region just above the spectral peak and then the slope changes to -5 beyond a certain multiple of the spectral peak frequency (e.g., Forristall, 1981; Kitaigorodskii, 1983; Hwang and Wang, 2001; Romero and Melville, 2010; Romero et al., 2012; Lenain and Melville, 2017; and references therein); or (b) maintaining the -4 equilibrium spectral slope with the introduction of a threshold wind speed for the reference scaling wind (e.g., Resio et al., 2004; Takagaki et al., 2018; and references therein).

Empirically, the spectral slope is determined from fitting the spectral components between $1.5f_p$ and $3f_p$ (Donelan et al., 1985) or $2f_p$ and $4f_p$ (Young 2006), where $f$ is the wave frequency and subscript $p$ indicates the spectral peak component. Numerical computations show that in the general ocean current conditions the Doppler frequency shift does not produce significant frequency change of these wave components in the neighborhood of the spectral peak used for spectral slope determination (Hwang et al., 2017). Extensive examination of the ocean surface wave spectra did not reveal convincing evidence of a constant spectral slope of -4, or -5, or -4 transitioning to -5. Instead, the measured spectral slope exhibits a stochastic random variation over a rather large range from milder than -3 to steeper than -6.

Similar random variation of the spectral slopes is also observed in the wavenumber spectra obtained from processing the 3D ocean surface topography. Hwang et al. (2000) report the average slope of azimuthally average wavenumber spectra varying over some range: -2.27±0.042 in one dataset and -2.17±0.087 in another; the ocean surface topography is measured by an airborne ocean surface topographic mapper similar to the one used in the research by Lenain and Melville (2017). The severe deviation of the observed spectral slopes from the expected equilibrium spectrum values indicates that wind waves in the ocean environment are generally in a state of nonequilibrium.

Random variation of the spectral slope is consistently observed in various sources of wave spectrum datasets. In particular, the fractions of observed spectral slopes within -4±0.25 and -5±0.25 are consistently found to be only about 20% to 25%. In essence, the constant spectral slope of -4, or -5, or -4 transitioning to -5 as stipulated in many theoretical equilibrium spectral models cannot be reproduced in field observations. In fact, similar information of spectral slope variability can be found in many published reports (e.g., Forrestal, 1981; Donelan et al., 1985; Young, 1998, 2006; Hwang et al., 2000, 2017); this issue will be further discussed in Sec. 2.1.

A correct determination of the spectral slope is fundamental to the accurate description of the wave spectrum, which prescribes the wave energy or momentum or slope distribution in different length or time scales. The wavelengths of special interest to this study are in the centimeter to decameter range. Since these waves cannot be described by the equilibrium spectrum as concluded from the observed significant spectral slope variation, the term "cmDm range" is adopted in this paper instead of the "equilibrium range".

Ocean surface low-pass-filtered mean square slope (LPMSS) measurements represent a very useful data source for the spectral slope investigation. The first systematic LPMSS dataset is reported by Cox and Munk (1954). By discharging a mixture containing crank case oil, diesel oil, and fish oil, they produce a large area covered with contiguous artificial slick that suppresses short waves, which are estimated to be less than about 0.3 m in wavelength. The technique works well for them to obtain LPMSS data in wind speeds up to about 10 m s$^{-1}$. More recently, the Global Navigation Satellite System Reflectometry (GNSSR) has generated considerable interest for its application in tropical cyclone (TC) monitoring (e.g., Ruf et al. 2016; and references therein). The achievement is culminated from many GNSSR LPMSS measurements extending the wind speed coverage to about 60 m s$^{-1}$ (e.g., Katzberg and Dunion, 2009; Gleason, 2013; Katzberg et al., 2013). The initiative also prompted new LPMSS measurements in hurricanes (e.g., Gleason et al., 2018). The low-passed cutoff wavelength of the L band (1.4 GHz) GNSSR technique is about 0.6 m.

This paper presents the spectral slope analysis using the traditional ocean buoy data and the LPMSS measurements. Sec. 2 describes the spectral slope observations from the two data sources. Sec. 3 discusses the relevant issues of the spectral coefficient and wind speed dependence, and Sec. 4 summarizes several conclusions.

## 2. Spectral slope analysis

### 2.1. In situ observations

Although there are many published reports of wave spectral measurements conforming to the ideal constant spectral slope of -4, or -5, or -4 transitioning to -5, close examinations of the field spectra reveal that the measured spectral slopes vary over some range. For example, Forristall (1981) reports saturation range spectral analysis of waves measured by wave staff and waverider buoy in the Gulf of Mexico, many of the wave spectral examples displayed in the paper deviate obviously from the ideal -4 or -5 or -4 transitioning to -5 spectral function; Donelan et al. (1985) mention a range of -3.5 to -5 spectral slopes in their wave spectra measured by a 14-element wave gauge array in Lake Ontario, Canada; Young (1998, 2006) shows a wide range of spectral slopes between -1.7 and -7 from analyzing more than 10 years ocean buoy directional wave spectra produced by TC wind fields in northwest coast of Australia; and Hwang et al. (2017) report a similar range of spectral slope variation in the wind wave spectra measured by wire gauges on a spar buoy in the Gulf of Tehuantepec, Mexico. As mentioned in Sec. 1, variable spectral slopes are also found in the wavenumber spectra analyzed with the 3D ocean surface topography (Hwang et al., 2000).

Fig. 1 shows four typical examples of the wave frequency spectra measured in the Central Bering Sea in 2006 at the National Data Buoy Center (NDBC) station 46035, the horizontal axis is the normalized frequency $f/f_p$. For reference, two line segments with slopes -4 and -5 are superimposed. For each spectrum, the wind speed $U_{10}$, the spectral peak wave period $T_p$, the dimensionless spectral peak frequency $\omega_{if}=\omega_p U_{10} g^{-1}=U_{10}/c_p$, and the magnitude of the frequency spectral slope $s_f$ are given in the square brackets in the legend; $g$ is the gravitational acceleration, $c_p$ is the phase speed of the spectral peak wave component, and $\omega=2\pi f$ is the angular frequency. The spectral slope is processed by least squares fitting the spectral components between $2\omega_p$ and $4\omega_p$, the range is marked by two vertical black dashed lines in the figure. It requires a leap of faith to accept that these observed wave spectra fit the mold of a spectral function with -4, or -5, or -4 transitioning to -5 slopes.

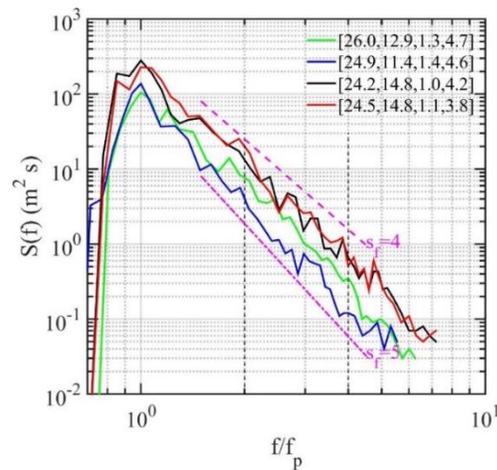

Fig. 1. Sample ocean wave spectra measured at the NDBC station 46035 in Central Bering Sea, given with the frequency normalized by the spectral peak frequency $S(f/f_p)$. The slopes of the two short magenta segments are -4 and -5; the two vertical black dashed lines represent the region for the spectral slope evaluation ($f/f_p$ between 2 and 4). The wind speed $U_{10}$, the peak wave period $T_p$, the inverse wave age $\omega_{if}$, and the magnitude of the spectral slope $s_f$ for each spectrum are listed in the legend in square brackets.

[Note: Waves in the ocean are generally made up of wind sea and swells. The swells arrive at the measurement station from distance storms. Only the wind sea properties can be related to the local wind condition. It is emphasized that all the spectra discussed in this paper are the wind sea portion of the measured spectra. The wind sea and swell separation uses the spectrum integration method described in Hwang et al. (2012).]

Fig. 2a shows the probability density function (pdf) $p(s_f)$ of the observed frequency spectral slopes from the full year (2006) data at the NDBC station 46035. The figure shows the results processed for three wind speed subgroups: 7~10, 10~15, and 15+ m s$^{-1}$. Only cases with $2 \leq s_f \leq 7$ and $\omega_\# \geq 0.8$ are included in this study. The number of spectra, and the mean and one standard deviation of the spectral slopes are given in parentheses in the legend. The normal distribution curves computed with the means and standard deviations are shown with solid lines of the same colors corresponding to the three wind speed groups. The mean spectral slopes observed are much steeper than -4, being -4.69, -4.88, and -4.92 respectively for the three wind speed groups. Fig. 2b shows the corresponding cumulative distribution function (CDF) $P(s_f)$. It is clear that only a minority fraction of the observations can be classified as within some reasonable intervals of a constant spectral slope. For example, the cumulative probability is about 15% for slopes within -4±0.25 and 20~25% for slopes within -5±0.25. Similar results are observed in other NDBC data, and they are generally consistent with the previous reports of Young (1998, 2006) and Hwang et al. (2017).

Hwang et al. (2017) stress the stochastic random nature of the observed $s_f$, and state that their "attempts to correlate $s_f$ with various wind and wave parameters, including $U_{10}$, $H_s$, $T_p$, their dimensionless combinations $\omega_\#$ and $\eta_\#$, and several swell-sea ratios, did not yield concrete result." We can add the acceleration or deceleration rate of wind speed or wave parameters to that list of factors failing to show convincing correlation with the observed spectral slope. Figs. 2c and 2d give examples of such unsuccessful correlation attempts. The dependent variable in Fig. 2c is $\omega_\#$ and the color code of the plotting symbol is wind sea significant wave height $H_{sw}$; the dependent variable in Fig. 2d is $U_{10}$ and the color code of the plotting symbol is $\omega_\#$ . [In the above quote from Hwang et al. (2017), $H_s$ is significant wave height, $\eta_\# = g^2 \eta_{rms}^2 U_{10}^{-4}$ is the dimensionless wave energy, and $\eta_{rms}^2$ is the variance of surface elevation.]

Fig. 2. The probability functions and examples showing the stochastic randomness of $s_f$. (a) The pdf of $s_f$

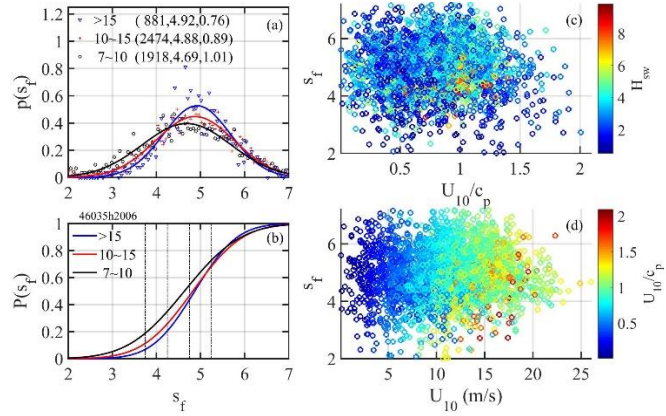

for three wind speed groups: $7 - 10$ m s$^{-1}$, $10 - 15$ m s$^{-1}$, and higher than 15 m s$^{-1}$. The number of spectra, and the mean and standard deviation of the spectral slopes in each wind speed group are given in parentheses in the legend; the corresponding Gaussian curves are superimposed with the same marker colors of the three groups. (b) The CDF of $s_f$ for the three wind speed groups shown in (a). (c) The scatter plot of $s_f$ against $U_{10}/c_p$ color coded with the wind sea significant wave height $H_{sw}$. (d) The scatter plot against $U_{10}$ color coded with $U_{10}/c_p$.

## 2.2. Remote sensing observations

### 2.2.1. Elevation and slope spectra

The spectral density in the frequency spectrum decreases sharply from the peak toward the high frequency region, for example, the spectral densities in the $2f_p$ to $4f_p$ region in the Fig. 1a samples are one to three orders of magnitude smaller than their peak values. Issues of spectral leakage and instrument or environmental noise are always a concern in the study of short scale waves using the wave elevation spectrum.

A different data source especially useful for the short wave analysis is the wave slope measurements. In the spectral representation $S_{slope}(k)=k^2 S(k)$ or $S_{slope}(\omega)=k^2 S(\omega)$, the $k^2$ weighting amplifies the short waves considerably; $k$ is wavenumber. Figs. 3a and 3b show an example comparing the elevation and slope frequency spectra computed with the G (general) spectrum model that accommodates a variable spectral slope (Hwang et al. 2017):

$$S_G(\omega) = \alpha g^2 \omega_p^{-5} \varsigma^{-s_f} \exp\left[-\frac{s_f}{4}\varsigma^{-4}\right]\gamma^\Gamma; \ \Gamma = \exp\left[-\frac{(1-\varsigma)^2}{2\sigma^2}\right]; \varsigma = \frac{\omega}{\omega_p}. \qquad (1)$$

The associated spectral parameters $\alpha$, $\gamma$, and $\sigma$ vary with $s_f$ and $\omega_{\#}$, the full expressions are given in Hwang et al. (2017) and summarized in the Appendix. [Young (2006) also gives the same spectral function but his spectral parameters do not vary with the spectral slope.]

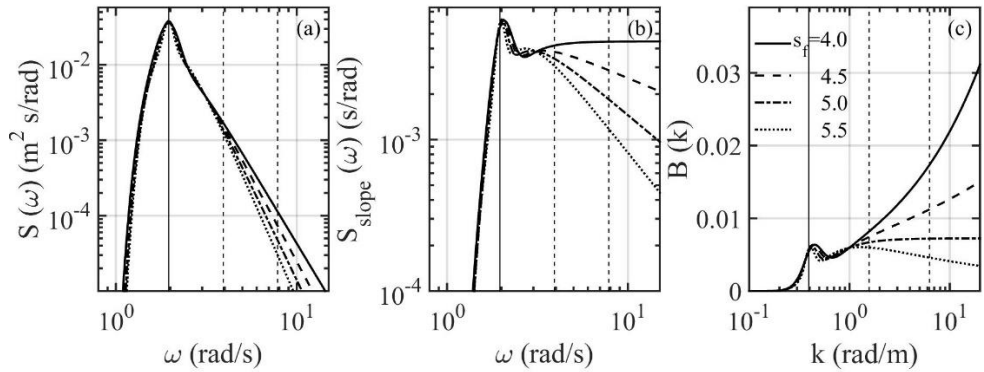

Fig. 3. Examples of the spectral slope effects on the wave spectrum. (a) The frequency elevation spectrum $S(\omega)$. (b) The frequency slope spectrum $S_{slope}(\omega)$. (c) The dimensionless wavenumber spectrum $B(k)$. Illustrated in each panel are the G spectrum model computed for four frequency spectral slopes $-s_f$=-4.0, -4.5, -5.0, and -5.5, $U_{10}$=10 m s$^{-1}$, and $\omega_{\#}$=2.

For Fig. 3, the wind speed $U_{10}$ is 10 m s$^{-1}$, the dimensionless spectral peak frequency $\omega_{\#}$ is 2, results for $s_f$ = 4.0, 4.5, 5.0, and 5.5 are illustrated; a vertical thin line marks the spectral peak frequency $\omega_p$, and two dashed lines mark $2\omega_p$ and $4\omega_p$. With the elevation spectrum, the signal in the region of interest ($2\omega_p$ to $4\omega_p$) for the spectral slope analysis is about one to three orders of magnitude smaller than the overall signal level represented by the spectral peak value (Fig. 3a). With the slope spectrum, they are of the same order of magnitude (Fig. 3b). The comparison highlights the advantage of the surface slope data over the elevation data for the study of the spectral slope.

The slope spectrum is frequently given as a function of wavenumber, especially in remote sensing applications for which the lateral length scale of the surface roughness is of great significance when considering resonance or tilting roughness components. An informative representation is the semilogarithmic plot of the dimensionless wavenumber spectrum $B(k)=k^3S(k)$; Fig. 3c shows the $B(k)$ corresponding to the frequency spectra shown in Figs. 3a and 3b. In this semilogarithmic plot, the areas under the curves between some wavenumber range are the mean square slopes integrated over the given wavenumber range. In contrast to the relatively minor impacts near the spectral peak region, the spectral slope effects are substantial in the cmDm range. This is especially important in the analysis of the spectral slope information using the surface slope data or vice versa.

### 2.2.2. LPMSS measurements

As mentioned in the Introduction, many surface slope data have been collected in support of remote sensing applications. Fig. 4 shows the wind speed dependency of the LPMSS data assembled from Cox and Munk (1954), Katzberg and Dunion (2009), Gleason (2013), Katzberg et al. (2013), and Gleason et al. (2018). In the figure legend, C54 is Cox and Munk (1954); K0913 combines Katzberg and Dunion (2009) and Katzberg et al. (2013), and G1318 combines Gleason (2013) and Gleason et al. (2018). Most of the K0913 and G1318 data are obtained inside tropical cyclones. More details about these LPMSS datasets have been given in the original papers and summarized in Section II of Hwang and Fan (2018), they are not repeated here.

Both methods of deriving the LPMSS from the sun glitter (C54) and GNSSR (K0913 and G1318) analyses are based on bistatic reflectometry, for which the specular reflection is the dominant mechanism and the tilting slopes are the surface roughness detected by the measurement systems. For the sun glint analysis of C54, the Sun is the electromagnetic (EM) source and the camera carried in the aircraft is the passive receiver. The suppression of the short scale waves for the low-passed filtering of the surface roughness is through the application of oil slicks. In K0913 and G1318, the EM source is the transmitters in the GNSS and the passive receivers on aircraft or satellite complete the measurement system. The low-passed filtering of the surface roughness is through the EM wavelength: the shortest wavelength of the low-passed tilting roughness is about three times the EM wavelength. For the GNSS L band frequency (1.41 GHz), the corresponding tilting roughness cutoff wavenumber $k_{tilt}$ is 9.85 rad m$^{-1}$, in the subsequent computation of LPMSS using the wave spectrum models, $k_{tilt}$ is rounded up to 10 rad m$^{-1}$.

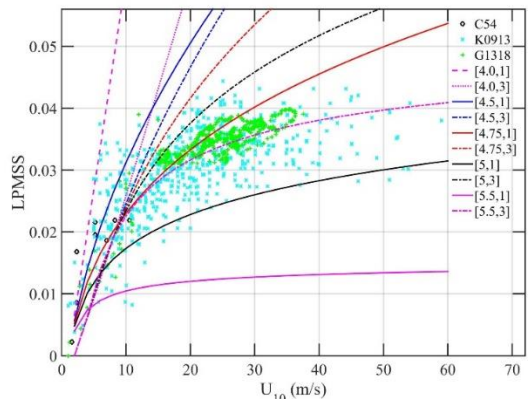

Fig. 4. Comparison of the LPMSS calculated with the G spectrum model with different wave ages and spectral slopes, superimposed on the background are the LPMSS measurements obtained by GNSSR (K0913 and G1318) and sun glitter analysis in slicked ocean surfaces (C54). The two numbers in the square brackets in the legend are $s_f$ and $\omega_{\#}$.

Superimposed in Fig. 4 are five sets of the G spectrum LPMSS calculation. The frequency spectral slopes -$s_f$ for the five sets are -4.0, -4.5, -4.75, -5.0, and -5.5. Each set contains two curves corresponding to $\omega_{\#} = 1$ and 3, which represent the approximate envelopes of the $\omega_{\#}$ range of the wind seas generated by hurricane wind fields as observed in hurricane reconnaissance and research missions (e.g., Hwang and Fan, 2017; and references therein). Several conclusions can be reached from this figure:

(a) In low and moderate winds ($U_{10}$ less than about 10 m s$^{-1}$), the LPMSS data show only a weak sensitivity to the spectral slope.

(b) As wind speed increases, the spectral slope effects become very distinctive and extracting the spectral slope information from the LPMSS data becomes more feasible.

(c) Spectral slopes milder than -4.5 can be excluded from consideration in moderate to high wind conditions.

(d) The LPMSS values show some complex dependence on three factors: $U_{10}$, $\omega_{\#}$, and $s_f$.

### 2.2.3. Spectral slope estimation using the LPMSS data

The same three factors ($U_{10}$, $\omega_{\#}$, and $s$) discussed in Sec. 2.2.2 also uniquely define the G spectrum. Hwang and Fan (2018) describe a procedure to use the LPMSS data to estimate the average spectral slope for the G spectrum. As mentioned earlier, the high-wind LPMSS data are collected inside hurricanes. The procedure takes advantage of the analysis results showing that the wind and wave fields inside TCs possess several similarity relationships useful for establishing the wind wave spectrum at any location inside the TC. The detail is presented in Section III of Hwang and Fan (2018); the following gives a brief summary.

Based on the analyses of the wind and wave measurements from eleven hurricane reconnaissance and research missions, it is found that the hurricane maximum wind speed $U_{10m}$ and its radius $r_m$ are two of the most important factors in the hurricane wind and wave similarity relationships. The wind and wave parametric models described in Hwang and Fan (2018) are built on these two hurricane parameters and a couple more to be discussed below.

The wind speed parametric model is based on the modified Rankine vortex model (Holland, 1980; Holland et al., 2010). To account for the wind field asymmetry, the vortex wind speed is normalized by $U_{10m\phi}$, which is the maximum wind speed along a radial transect with a given azimuth angle $\phi$,

$$\frac{U_{10}(r,\phi)}{U_{10m\phi}} = \begin{cases} r_*, & r_* \leq 1 \\ r_*^{-0.5}, & r_* > 1 \end{cases},$$ (2)

where $r$ is the radial distance from the hurricane center, the normalized radial distance is $r_* = r/r_m$, the azimuth angle is measured from the hurricane heading, increasing counterclockwise (CCW). Based on the examination of archived NOAA HWIND 2D wind fields in several historical hurricanes, the following formula is used for $U_{10m\phi}$:

$$\frac{U_{10m\phi}}{U_{10m}} = 1 - a_{1U}\left[1 - \cos\left(\phi - \phi_m\right)\right],$$ (3)

where $\phi_m$ is the azimuth angle of the location of the overall maximum wind speed $U_{10m}$; the $U_{10m}$ location is mostly in the right-hand side relative to the hurricane heading, with higher probability in the right-front quadrant ($\phi$ between 270° and 360°) than the right-back quadrant ($\phi$ between 180° and 270°). The wind field asymmetry factor $a_{1U}$ varies mostly between about 0.1 and 0.2.

For the wave spectrum computation, the most useful characteristic wave parameter is $\omega_{\#}$. In the wave parametric model, it is given in the normalized form: $\omega_* = \omega_{\#}/U_{10m}$, the prominent feature of $\omega_*(\phi)$ is the sinusoidal variation:

$$\omega_* = a_0\left(r_*\right) + a_1\left(r_*\right)\cos\left[\phi + \delta\left(r_*\right)\right].$$ (4)

The coefficients $a_0$, $a_1$, and $\delta$ vary with $r_*$ and they are empirically determined from the hurricane reconnaissance and research measurements

$$a_0 = \begin{cases} 0.0056r_*, & r_* \leq 1 \\ 0.0056r_*^{-0.5}, & 1 < r_* \leq 3.3 \\ 0.03, & 3.3 < r_* \end{cases}.$$ (5)

$$\frac{a_1}{a_0} = \begin{cases} -0.1, & r_* \leq 1 \\ -0.1 r_*^{-1/3}, & 1 < r_* \leq 2 \\ -0.27, & 2 < r_* \end{cases} . \tag{6}$$

$$\delta = \begin{cases} -60 + 240(1 - r_*), & r_* \leq 1.1 \\ -80 r_*^{-1/3}, & 1.1 < r_* \end{cases} . \tag{7}$$

With $U_{10}(r,\phi)$ and $\omega_m(r,\phi)$ given by the parametric models, the spectral slope $-s_f(r,\phi)$ becomes the only remaining unknown factor in the G spectral function for the LPMSS computation at any location $(r,\phi)$ inside the TC. Fig. 5a shows two examples of the computation with $s_f$=4.7 and 5.0, superimposed in the figure are the LPMSS data. It is clear that neither constant slope produces satisfactory comparison with the full set of the LPMSS data.

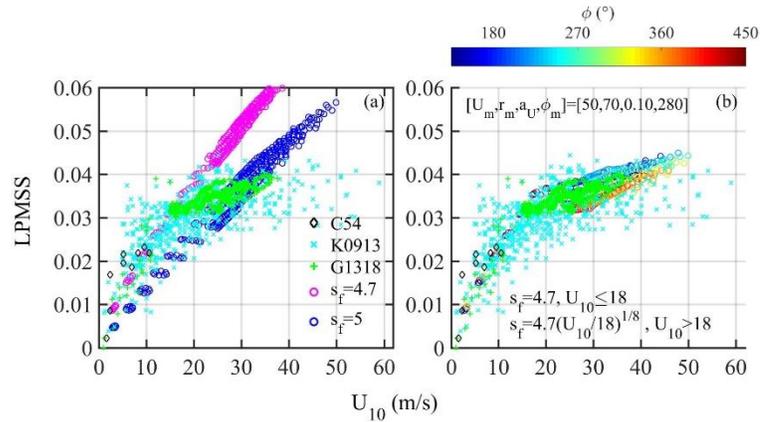

Fig. 5. Comparison of the modeled and measured LPMSS. (a) Results based on two constant $s_f$ values (4.7 and 5). (b) Results based on variable spectral slope given in (8). The hurricane parameters [$U_{10m}$, $r_m$, $a_{1U}$, $\phi_m$] are [50 m s$^{-1}$, 70 km, 0.1, 280°]. In (b) the circle is color-coded with the azimuth angle $\phi$ from the hurricane heading.

Through numerical experimentation matching the modeled LPMSS with field measurements, Hwang and Fan (2018) recommend the following wind speed function for the average $s_f$

$$s_f = \begin{cases} s_1, & U_{10} \leq U_1 \\ s_1 (U_{10}/U_1)^q, & U_{10} > U_1 \end{cases} . \tag{8}$$

Fig. 5b shows the significantly improved model-measurement comparison with the parameter values $[s_1, U_1, q] = [4.7, 18 \text{ m s}^{-1}, 1/8]$; plug those numbers into (8) the average $s_f$ increases gradually from 4.70 to 5.66 for $U_{10}$ from 18 to 80 m s$^{-1}$. The simulated data reveal some minor azimuthal dependence as illustrated by the color code of the plotting symbol. The simulated wind and wave fields in Fig. 5 are generated by the wind and wave parametric models described earlier with the hurricane parameters $[U_{10m}, r_m, a_{1U}, \phi_m] = [50 \text{ m s}^{-1}, 70 \text{ km}, 0.1, 280°]$. More extensive discussions and model-measurement comparisons with a wide range of hurricane parameters are given in Hwang and Fan (2018).

## 3. Discussion

### 3.1. Spectral coefficient

For a power function $S(\omega) = K_f \omega^{-s_f}$ or $S(k) = K_k k^{-s_k}$, there is a close correlation between the proportionality coefficient and the exponent from least squares fitting of experimental data. The dimensionless spectral coefficient is related to $K_f$ or $K_k$. For example, the high frequency portion of the G spectrum model can be written as

$$S(\omega) = \alpha_f g^2 \omega_p^{s_f - 5} \omega^{-s_f} .$$ (9)

Applying $S(\omega) d\omega = S(k) dk$, the corresponding wavenumber spectrum is

$$S(k) = \alpha_k k_p^{s_k - 3} k^{-s_k} ,$$ (10)

where $\alpha_k = \alpha_f/2$ and the wavenumber spectral slope is $-s_k = -(s_f + 1)/2$; the dimensionless spectral coefficient can be calculated from the fitted proportionality coefficient, i.e., $\alpha_f = K_f \left( g^2 \omega_p^{s_f - 5} \right)^{-1}$ or equivalently $\alpha_f = K_f \left( g^{s_f - 3} \omega_\#^{s_f - 5} U_{10}^{5 - s_f} \right)^{-1} .$

Fig. 6 shows an example of data processing to obtain the spectral coefficients of the NDBC buoy spectra used for illustration in this study (Figs. 1 and 2). The top row shows the results of data processing based on the measured wave spectral slope, and the bottom row shows the results based on assuming a fixed -4 frequency spectral slope. To make the distinction, subscripts $s$ and 4 are appended to the coefficient labels $K_f$ and $\alpha_f$ in the upper and lower rows, respectively. It is clear that assuming a constant spectral slope suppresses the ranges of the processed coefficients; the computational results are presented in log-log scales in the left column and linear scales in the right column with the same horizontal and vertical spans for easy comparison.

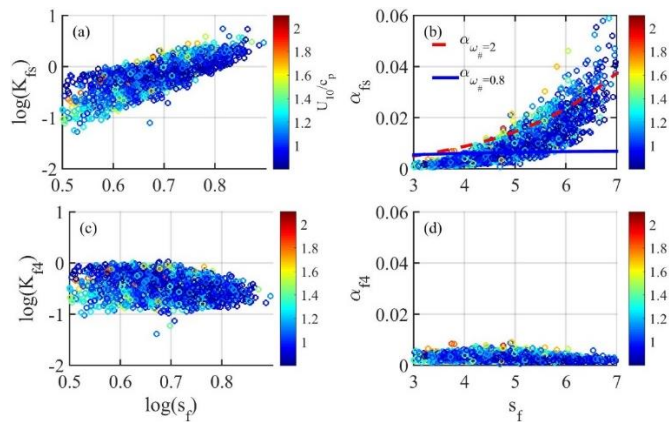

Fig. 6. The proportionality coefficient $K_f$ of the power function fitted to the frequency range of $2\omega_p - 4\omega_p$ of the wave frequency spectrum. (a) $K_{fs}$: processed with the observed spectral slope for the exponent of the fitting power function, and (c) $K_{f4}$: processed with -4 as the exponent of the fitting power function. (b) and (d) are the corresponding spectral coefficient $\alpha_f$; the color coding is $\omega_{ff}=U_{10}/c_p$. In (b) the two smooth curves are computed G spectrum $\alpha_f$ for $\omega_{ff}=0.8$ and 2.0 (Hwang et al., 2017).

The G spectral coefficient $\alpha_f$ is a function of both $s_f$ and $\omega_{ff}$ (Appendix), the blue solid and red dashed lines in Fig. 6b are computed with $\omega_{ff}=0.8$ and 2, respectively. The $\alpha_f$ dependence on $s_f$ is weak at mature stage and much stronger in younger seas. The field results are rather scattered and yield only a crude agreement.

*3.2. Wind speed dependence*

### 3.2.1. Frequency wave spectrum analysis

To examine the wind speed dependency of the spectral function in the cmDm range, (9) can be rewritten as

$$S(\omega) = \left[ \alpha_f g^{s_f - 3} \omega_{\#}^{s_f - 5} \right] U_{10}^X \omega^{-s_f} = K_f \omega^{-s_f} , \qquad (11)$$

Where $X = 5 - s_f$ is the explicit wind speed exponent. Because the product enclosed in the square brackets in (11) is a function of $s_f$ and $\omega_{\#}$, the latter contains $U_{10}$, i.e., $\omega_{\#} = \omega_p U_{10} g^{-1}$, and the mean value of $s_f$ shows some $U_{10}$ dependence, it is necessary to sort the data into small $s_f$ and $\omega_{\#}$ bins in order to extract the explicit wind speed dependence embedded in $K_f$. Fig. 7a shows an example displaying $K_f$ within the $\omega_{\#}$ bin 1.1~1.3 with different color markers for the four $s_f$ bins identified in the legend. The power function exponent obtained from least squares fitting are noted at the left end of each fitted curve. In Fig. 7b, the exponents, here denoted as $X_K$, as a function of the average $s_f$ are illustrated with black markers for four different $\omega_{\#}$ bins identified in the legend. If $\alpha_f$ is independent on $U_{10}$ aside from the $\omega_{\#}$ and $s_f$ dependencies as given in the formulation (Appendix), $X_K$ is expected to be the same as $X = 5 - s_f$, which is illustrated with a dashed line in the figure. The data scatter is rather large but the values of $X_K$ are closer to $6 - s_f$, which is shown with a dashed-dotted line. The result suggests that $\alpha_f$ has additional quasi-linear wind speed dependence. The least squares fitting is also applied to $\alpha_f$, the resulting wind speed exponents, here denoted as $X_\alpha$, are shown in Fig. 7b in red color with the same markers for the four different $\omega_{\#}$ bins; the $X_\alpha$ values scatter about $1 \pm 0.4$.

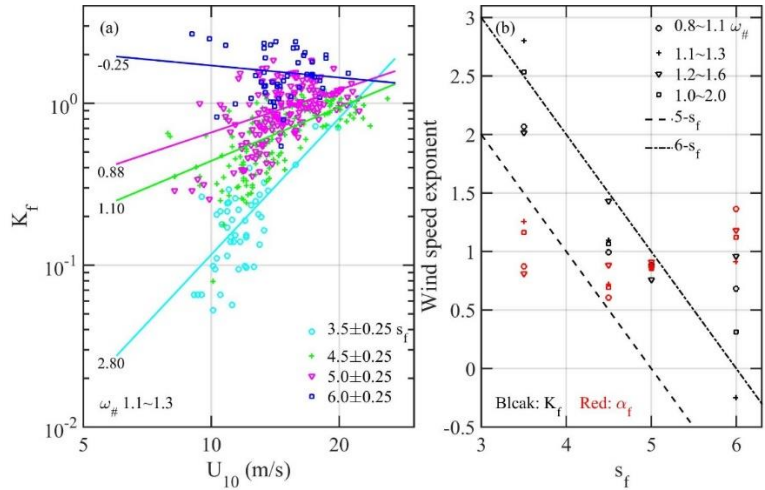

Fig. 7. (a) Empirical determination of the wind speed exponent in the coefficient $K_{fU}$ of the frequency spectrum. The data for four $s_f$ bins identified in the legend and within the $\omega_{\#}$ bin 1.1~1.3 are used for illustration. The least squares fitted curves are shown with the same colors of the four groups; the exponent for each curve is labeled at the left end of the curve. (b) The wind speed exponents of $K_f$ and $\alpha_f$ processed from four $\omega_{\#}$ bins identified in the legend, dashed and dashed-dotted lines show 5-$s_f$ and 6-$s_f$, respectively.

### 3.2.2. Wavenumber observations

The wind sensitivity is of great interest to ocean remote sensing and there is a big volume of research results on the subject, especially with respect to the exponent of the power function relating the measured EM signal (e.g., the radar backscattering cross section or microwave radiometer brightness temperature) to the wind speed or wind friction velocity. Fig. 8a shows a summary of results assembled from an extensive collection of remote sensing and in situ measurements of the wind friction velocity exponent of short scale waves. More details of the analyses are summarized in Hwang (1997), Trokhimovski and Irisov (2000), and Hwang and Wang (2004); the three references are labeled H97, T00, and H04, respectively in the legend. Of special interest to this study is the wavenumber range between about 1 and 20 rad m$^{-1}$. Many interesting surface features such as the breaking patch sizes fall into the corresponding spectral wavelengths between about 0.3 and 6 m (e.g., Hwang et al., 1989; Ding and Farmer, 1994; Frasier et al., 1998; Phillips et al., 2001; Melville and Matusov, 2002; Hwang and Wang, 2004; Hwang, 2007; Hwang et al., 2008, 2016; and references therein). As shown in Fig. 8a, for these wavelengths the wind friction velocity exponent is much less than one so the wind sensitivity clearly deviates from the linear dependence expected from the equilibrium spectrum function (e.g., Phillips, 1985)

$$S_e\left(k\right) = b_e \frac{u_*}{c} k^{-3} = b_e u_* g^{-0.5} k^{-2.5}. \tag{12}$$

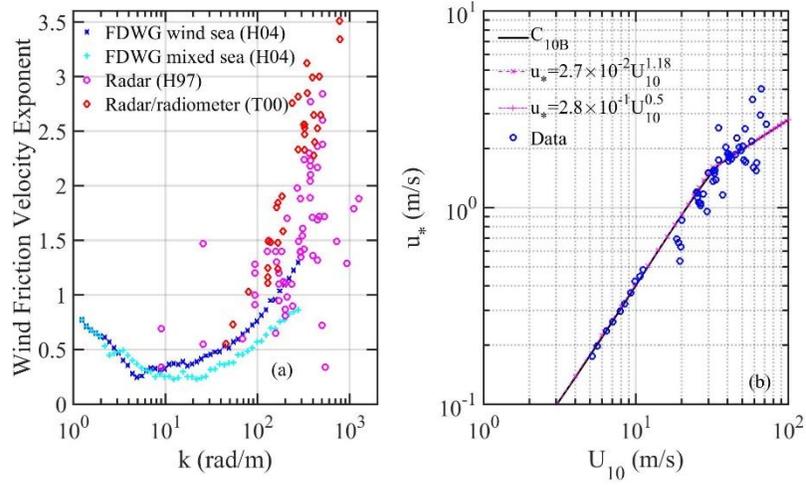

Fig. 8. (a) The wind friction velocity exponent obtained from wind sensitivity analysis of remote sensing and in situ short wave data, more details are given in Hwang (1997), Trokhimovski and Irisov (2000), and Hwang and Wang (2004). (b) Power function approximation of $u_*(U_{10})$ in two wind speed ranges, the field data and the drag coefficient function $C_{10B}$ are described in Hwang (2018).

The observed deviation from the explicit linear $u_*$ dependence is again indicative of the nonequilibrium nature of the cmDm waves in the ocean environment. From dimensional consideration, a general wavenumber spectrum function can be written as

$$S\left(k\right) = b_s\left(\frac{u_*}{c}\right)^Y k^{-3} = b_s u_*^Y g^{-0.5Y} k^{-s_k}, \tag{13}$$

where the magnitude of the wavenumber spectral slope is $s_k$=3-0.5$Y$, and $Y$ is the exponent of the explicit wind friction velocity power function. Because $s_k$=($s_f$+1)/2, the explicit wind exponent is related to the wavenumber or frequency spectral slope as

$$Y = 6 - 2s_k = 5 - s_f = X .$$

(14)

The analysis leading to (14) yields the expectation of the same wind dependence whether expressed as the wind speed or wind friction velocity power function. However, because of the nonlinear drag coefficient dependence on wind speed, the relation between $u_*$ and $U_{10}$ is approximately $u_* \approx 2.7 \times 10^{-2} U_{10}^{1.18}$ for $U_{10}$ less than about 30 m s$^{-1}$ (Fig. 8b); the data shown in Fig. 8a are all collected in conditions with $U_{10}$<30 m s$^{-1}$; see H97, T00, and H04 for more detail. As a result $X$=1.18$Y$ is expected. The experimental data scatter is rather large (e.g., see Figs. 7 and 8) for detecting the 18% difference between the $u_*$ and $U_{10}$ dependencies. Interestingly, in very high winds $X$=0.5$Y$ is expected from $u_* \approx 2.8 \times 10^{-1} U_{10}^{0.5}$ for $U_{10}$ greater than about 30 m s$^{-1}$ (Fig. 8b; Hwang, 2018).

## 4. Conclusions

The wave spectral slopes in field measurements vary over a wide range and deviate from the constant values expected in the theoretical equilibrium spectrum models. This observation indicates that the ocean surface waves are generally in a state of nonequilibrium.

Expressed as a power function for the short (cmDm) wave spectrum $S(\omega) = K_f \omega^{-s_f}$ or $S(k) = K_k k^{-s_k}$, the proportionality coefficient and wind sensitivity of the spectral function is closely related to the spectral slope. Processing of the spectral measurements assuming a constant spectral slope severely distorts the true nature of the wind wave spectral properties. The correct determination of the spectral slope is critical to an accurate prescription of the spectral properties of cmDm waves.

**Appendix. The G (general) wind wave spectrum model**

The G model is given as (Hwang et al., 2017):

$$S(\omega) = \alpha g^2 \omega_p^{-5} \varsigma^{-s_f} \exp\left[-\frac{s_f}{4}\varsigma^{-4}\right] \gamma^{\Gamma}; \; \Gamma = \exp\left[-\frac{(1-\varsigma)^2}{2\sigma^2}\right]; \varsigma = \frac{\omega}{\omega_p}. \qquad (A1)$$

The associated spectral parameters vary with $s_f$ and $\omega_\#$:

$$\begin{aligned} \alpha &= \alpha_1\left[1-0.3\tanh(0.1\omega_\#)\right] \\ \gamma &= \gamma_1\left[1-0.5\tanh(0.1\omega_\#)\right], \\ \sigma &= A_\sigma + a_\sigma \log(\omega_\#) \end{aligned} \qquad (A2)$$

where

$$\begin{aligned} \alpha_1 &= A_\alpha \omega_\#^{a_\alpha} \\ \gamma_1 &= A_\gamma + a_\gamma \log(\omega_\#) \end{aligned} \qquad (A3)$$

The coefficients $A_\alpha, a_\alpha, A_\gamma, a_\gamma, A_\sigma,$ and $a_\sigma$ are functions of $s_f$:

$$\begin{aligned} A_\alpha &= 1.30\times10^{-3}s_f + 1.64\times10^{-3}, \\ a_\alpha &= 4.83\times10^{-1}s_f - 1.49, \\ A_\gamma &= 4.42\times10^{-1}s_f + 3.93\times10^{-1}, \\ a_\gamma &= -3.63s_f + 19.74, \\ A_\sigma &= -5.39\times10^{-2}s_f + 3.44\times10^{-1}, \\ a_\sigma &= 2.05\times10^{-9}s_f + 5.5\times10^{-2}. \end{aligned} \qquad (A4)$$

**Acknowledgements**

data are archived at http://www.rms.com/perils/hwind/legacy-archive/. The processing codes and data

segments can also be obtained by contacting the corresponding author.

**References**

Cox, C. S., and Munk, W., 1954. Statistics of the sea surface derived from sun glitter. J. Mar. Res., 13, 198-227.

Ding, L., and Farmer, D. M., 1994. Observations of breaking surface wave statistics. J. Phys. Oceanogr., 24, 1368-1387.

Donelan, M. A., Hamilton, J., and Hui, W. H., 1985. Directional spectra of wind-generated waves, Phil. Trans. Roy. Soc. Lond., A315, 509-562.

Forristall, G. Z., 1981. Measurements of a saturated range in ocean wave spectra. J. Geophys. Res., 86, 8075–8084, doi:10.1029/JC086iC09p08075.

Frasier, S. J., Liu, Y., and McIntosh, R. E., 1998. Space-time properties of radar sea spikes and their relation to wind and wave conditions. J. Geophys. Res., 103, 18745-18757.

Gleason, S., 2013. Space based GNSS scatterometry: Ocean wind sensing using empirically calibrated model. IEEE Trans. Geosci. Remote Sens., 51, 4853-4863.

Gleason, S., Zavorotny, V., Hrbek, S., PopStefanija, I., Walsh, E., Masters, D., and Grant, M., 2018. Study of surface wind and mean square slope correlation in hurricane Ike with multiple sensors. IEEE J. Sel. Topics Appl. Earth Observ. in Remote Sens., 11, 1975-1988.

Hasselmann, K. et al., 1973. Measurements of wind-wave growth and swell decay during the Joint North Sea Wave Project (JONSWAP). Deutsch. Hydrogr. Z., Suppl. A8(12), 95pp.

Hasselmann, K., Ross, D. B., Müller, P., and Sell, W., 1976. A parametric wave prediction model. J. Phys. Oceanogr., 6, 200-228.

Holland, G. J., 1980. An analytic model of the wind and pressure profiles in hurricane. Mon. Wea. Rev., 108, 1212-1218.

Holland, G. J., Belanger, J. I., and Fritz, A., 2010. A revised model for radial profiles of hurricane winds.

Mon. Wea. Rev., 136, 4393-4401.

Hwang, P. A., 1997. A study of the wavenumber spectra of short water waves in the ocean. Part 2: Spectral model and mean square slope. J. Atmos. and Oceanic Tech., 14, 1174-1186.

Hwang, P. A., 2005. Wave number spectrum and mean square slope of intermediate-scale ocean surface waves. J. Geophys. Res., 110, C10029.

Hwang, P. A., 2007. Spectral signature of wave breaking in surface wave components of intermediate length scale. J. Mar. Sys., 66, 28-37, doi:10.1016/j.jmarsys.2005.11.015.

Hwang, P. A., 2018. High wind drag coefficient and whitecap coverage derived from microwave radiometer observations in tropical cyclones. J. Phys. Oceanogr., 48, 2221-2232, doi: 10.1175/JPO-D-18-0107.1.

Hwang, P. A., and Fan, Y., 2017. Effective fetch and duration of tropical cyclone wind fields estimated from simultaneous wind and wave measurements: Surface wave and air-sea exchange computation. J. Phys. Oceanogr., 47, 447-470, doi: 10.1175/JPO-D-16-0180.1.

Hwang, P. A., and Fan, Y., 2018. Low-frequency mean square slopes and dominant wave spectral properties: Toward tropical cyclone remote sensing. IEEE Trans. Geos. Rem. Sens., 56, 7359-7368, doi: 10.1109/TGRS.2018.2850969.

Hwang, P. A, and Wang, D. W., 2001. Directional distributions and mean square slopes in the equilibrium and saturation ranges of the wave spectrum. J. Phys. Oceanogr., 31, 1346-1360.

Hwang, P. A., and Wang, D. W., 2004. An empirical investigation of source term balance of small scale surface waves. Geophys. Res. Lett., 31, L15301, 1-5, doi:10.1029/2004GL020080.

Hwang, P. A., Xu, D., and Wu, J., 1989. Breaking of wind-generated waves: Measurements and characteristics. J. Fluid Mech., 202, 177-200.

Hwang, P. A., Wang, D. W., Walsh, E. J., Krabill, W. B., and Swift, R. N. 2000. Airborne measurements of the directional wavenumber spectra of ocean surface waves. Part 1. Spectral slope and dimensionless spectral coefficient. J. Phys. Oceanogr., 30, 2753-2767.

Hwang, P. A., Sletten, M. A., and Toporkov, J. V., 2008. Analysis of radar sea return for breaking wave investigation. J. Geophys. Res., 113, C02003, 1-16, doi:10.1029/2007JC004319.

Hwang, P. A., Ocampo-Torres, F. J., and García-Nava, H., 2012. Wind sea and swell separation of 1D wave spectrum by a spectrum integration method. J. Atmos. Oceanic Tech., 29, 116-128, doi: 10.1175/JTECH-D-11-00075.1.

Hwang, P. A., Savelyev, I. B., and Anguelova, M. D., 2016. Breaking waves and near-surface sea spray aerosol dependence on changing winds. Wave breaking efficiency and bubble-related air-sea interaction processes. Proc. 7th Symp. Gas Transfer at Water Surfaces, IOP Conf. Series: Earth and Environmental Science 35 (2016) 012004 doi:10.1088/1755-1315/35/1/012004, 1-9.

Hwang, P. A., Fan, Y., Ocampo-Torres, F. J., and García-Nava, H., 2017. Ocean surface wave spectra inside tropical cyclones. J. Phys. Oceanogr., 47, 2293-2417, doi: 10.1175/JPO-D-17-0066.1.

Katzberg, S. J. and Dunion, J., 2009. Comparison of reflected GPS wind speed retrievals with dropsondes in tropical cyclones. Geophys. Res. Lett., 36, L17602, doi:10.1029/2009GL039512.

Katzberg, S. J., Dunion, J., and Ganoe, G. G., 2013. The use of reflected GPS signals to retrieve ocean surface wind speeds in tropical cyclones. Radio Sci., 48, 371–387, doi:10.1002/rds.20042.

Kitaigorodskii, A. 1983. On the theory of the equilibrium range in the spectrum of wind-generated gravity waves. J. Phys. Oceanogr., 13, 816-827.

Lenain, L., and Melville, W. K., 2017: Measurements of the directional spectrum across the equilibrium saturation ranges of wind-generated surface waves. J. Phys. Oceanogr., 47, 2123-2138.

Melville, W. K., and Matusov, P., 2002. Distribution of breaking waves at the sea surface. Nature, 417, 58-63.

Phillips, O. M., 1958. The equilibrium range in the spectrum of wind generated waves. J. Fluid Mech., 4, 785-790.

Phillips, O. M., 1985. Spectral and statistical properties of the equilibrium range in wind-generated gravity waves. J. Fluid Mech., 156, 505-531.

Phillips, O. M., Posner, F. L., and Hansen, J. P. 2001. High range resolution radar measurements of the speed distribution of breaking events in wind-generated ocean waves: Surface impulse and wave energy dissipation rates. J. Phys. Oceanogr., 31, 450-460.

Pierson, W. J., and Moskowitz, L., 1964. A proposed spectral form for fully developed wind seas based on the similarity theory of S. A. Kitaigorodskii. J. Geophys. Res., 69, 5181-5190.

Resio, D. T., Long, C. E., and Vincent, C. L., 2004. Equilibrium-range constant in wind-generated wave spectra. J. Geophys. Res., 109, C01018, doi:10.1029/2003JC001788.

Romero, L., and Melville, W. K., 2010. Airborne observations of fetch-limited waves in the gulf of Tehuantepec. J. Phys. Oceanogr., 40, 441-465.

Romero, L., Melville, W. K., and Kleiss, J. M., 2012. Spectral energy dissipation due to surface wave breaking. J. Phys. Oceanogr., 42, 1421–1444, doi:10.1175/JPO-D-11-072.1.

Ruf, C. S., Atlas, R., Chang, P. S., Clarizia, M. P., Garrison, J. L., Gleason, S. Katzberg, S. J., Jelenak, Z. Johnson, J. T., Majumdar, S. J., O'Brien, A., Posselt, D. J., Ridley, A. J., Rose, R. J., an Zavorotny, d V. U., 2016. New ocean winds satellite mission to probe hurricanes and tropical convection. Bull. Amer. Meteor. Soc., 385-395, doi:10.1175/BAMS-D- 14-00218.1.

Takagaki, N, Takane, K., Kumamaru, H., Suzuki, N., and Komori, S., 2018. Laboratory measurements of an equilibrium-range constant for wind waves at extremely high wind speeds. Dyna. Atmos. and Oceans, 84, 22-32, DOI: 10.1016/j.dynatrace.2018.08.003.

Trokhimovski, Y. G., and Irisov, V. G., 2000. The analysis of wind exponents retrieved from microwave radar and radiometric measurements. IEEE Trans. Geosci. Remote Sens., 38, 470-479.

Wright, J. W., 1966. Backscattering from capillary waves with application to sea clutter. IEEE Trans.

   Antennas Propag., 14, 749-754.

Wright, J. W., 1968. A new model for sea clutter. IEEE Trans. Antennas Propag., 16, 217-223.

Young, I. R., 1998. Observations of the spectra of hurricane generated waves. Ocean Eng., 25, 261-276.

Young, I. R., 2006. Directional spectra of hurricane wind waves. J. Geophys. Res., 111, C08020, 1-14.